\documentclass[12pt]{JHEP3}
\usepackage{amsmath,amssymb}
\usepackage{yfonts}
\usepackage{graphicx}
\usepackage{comment}
\usepackage[square,comma, sort&compress]{natbib}

\includecomment{setup}
\includecomment{graphresults}
\includecomment{conductivity}
\includecomment{appendices}
\includecomment{references}

\newcommand{\be}{\begin{equation}}
\newcommand{\ee}{\end{equation}}
\newcommand{\ben}{\begin{equation*}}
\newcommand{\een}{\end{equation*}}
\newcommand{\beqalabel}[1]{\begin{subequations}\label{#1}\begin{eqnarray}}
\newcommand{\beqa}{\begin{subequations}\begin{eqnarray}}
\newcommand{\eeqa}{\end{eqnarray}\end{subequations}}
\newcommand{\beqad}{\begin{eqnarray}}
\newcommand{\eeqad}{\end{eqnarray}}
\newcommand{\beqan}{\begin{eqnarray*}}
\newcommand{\eeqan}{\end{eqnarray*}}

\newcommand{\qn}{\textswab{q}}
\newcommand{\wn}{\textswab{w}}

\newcommand{\im}{\hbox{\,Im\,}}

\def\d{\partial}

\def\NN{\mathcal{N}}
\def\OO{\mathcal{O}}
\def\SS{\mathcal{S}_{DBI}}
\def\LSS{\tilde {\mathcal{S}}_{DBI}}
\def\dd{\mathrm{d}}
\def\current{\tilde \j_z}
\def\bardens{\tilde n_q}
\def\efield{ e}
\def\dlcond{\tilde \sigma}
\def\dA{{\cal A}}
\def\dB{{\cal B}}

\def\pperp{\perp\!\perp}

\title{Holographic Spectral Functions in Metallic AdS/CFT}

\author{ Javier Mas\footnote{jamas@fpaxp1.usc.es}, Jonathan P. Shock\footnote{shock@fpaxp1.usc.es} and Javier Tarr\'\i o\footnote{tarrio@fpaxp1.usc.es}
\\
 Departamento de F\'\i sica de Part\'\i culas,
Universidade de Santiago de Compostela \\ 
and\\
Instituto Galego de F\'\i sica de Altas Enerx\'\i as (IGFAE)\\

E-15782 Santiago
de Compostela, Spain\\
}

\preprint{ }

\abstract{
We study the holographic $D3/D7$ setup dual to ${ \cal N}=4$ supersymmetric Yang-Mills with quenched fundamental matter. 
We extend the previous analyses of conductivity and photoproduction to the case where there is  a finite electric field. Due to the electric field a special region in the $D7$-brane geometry, labelled the singular shell, appears generically, and the computation of correlators involves a careful study of the indicial exponents
both at this singular region and at the horizon. We show that there is a unique choice consistent with the known expression for the electrical conductivity found by Karch and O'Bannon [1]. We explore the parameter space spanned by the quark mass, the baryon density and the electric field. We find a region where the conductivity and photoproduction change rapidly and trace this behavior to competing effects which manifest themselves as a crossover behavior in the probe brane embeddings.
}

\begin{document}

\section{Introduction}
Since its original formulation, the AdS/CFT correspondence has been applied to an extremely diverse range of phenomena. Only recently, the applications to non-relativistic systems \cite{Son:2008ye} and superconductivity \cite{Ammon:2009fe,Hartnoll:2009sz} have been elucidated and a large number of papers on these subjects continue to be published. One of the most recent applications of the correspondence has been to study gauge theories in the presence of electric and magnetic fields \cite{Karch:2007pd,Filev:2007gb,Filev:2007qu,Albash:2007bk,Erdmenger:2007bn,Albash:2007bq,Filev:2009xp,Bergman:2008sg,Bergman:2008qv}, a useful way to explore the properties of many physical systems.

In this paper we are interested in the study of strongly coupled plasmas, in particular the $Dp/Dq$ brane intersections holographically dual to Yang-Mills theory with fundamental hypermultiplets \cite{hep-th/0205236,Erdmenger:2007cm}. We will use the benchmark $D3/D7$ system in order to study the properties of such plasmas, generalising wherever possible to the $Dp/Dq$ intersection.

One of the first additions to the most basic AdS/QGP (Anti-de-Sitter/Quark Gluon Plasma) calculations was the introduction of a finite charge density \cite{hep-th/0611021,hep-th/0611099}. This corresponds to studying a $Dq$-brane probe in a black $Dp$-brane background (a geometry with an AdS boundary and a black hole at the centre) with the time component of a world-volume gauge field on the $Dq$-brane turned on. One of the most striking consequences of the addition of finite charge density was the disappearance of the, so called, ``Minkowski embeddings" \cite{hep-th/0611099}. These Minkowski embeddings correspond to $Dq$-branes which wrap vanishing cycles which end away from the black hole horizon. These solutions are characterised by a discrete spectrum of excitations corresponding to infinitely long-lived bound states. However,  no matter how small the quark density introduced, the bulk gauge field flux lines need somewhere to end, and the only option is the horizon, removing the possibility of having Minkowski embeddings. An immediate consequence of this is that the introduction of heavy flavors with finite charge density leads to flavor probe brane embeddings which, as they run in from the AdS boundary into the IR, are flat all the way down to some small region where they develop a long spike which eventually hits the horizon. All embeddings in this setting then take the form of ``Black Hole", or ``Schwarzschild embeddings". In more physical  terms, this means that with any finite quark density, all mesons are unstable and the spectral function becomes continuous. It also means that the ``fundamental phase transition" between stable and unstable mesons disappears. 

In this setting, the magnitude of the charge density can be used as a control parameter in which the smooth, continuous spectral functions tend, in the vanishing charge density limit, to those of the stable phase with its signature discrete spectral function. In this way one can approach the stable meson phase arbitrarily close. In the limit where the spectral function takes the form of sharp peaks one can study the dispersion relation of the modes in the plasma by tracking the peak positions in the frequency plane as a function of their momenta. This was performed for transverse excitations in \cite{arXiv:0804.2168} and for the more involved longitudinal ones in \cite{arXiv:0805.2601}.
For the transverse case subluminal asymptotic velocities were observed, while in the second, some evidence for superluminal propagation was obtained. This second, rather more exotic behavior is currently under study by looking at the fully fledged quasinormal mode calculation \cite{wip}.

The next step in exploring these kinds of systems is the introduction of electric and/or magnetic fields within the holographic setups at zero and finite temperature. In \cite{Karch:2007pd,Albash:2007bk,Erdmenger:2007bn,Albash:2007bq,O'Bannon:2007in} the phase structure of these systems was studied in detail, while the spectrum of excitations was also looked at in a certain subsector of the phase diagram. In the present study we will be interested in extending this work to the rest of the phase diagram in the case of the electric field, where the behavior is more complicated than in the magnetic case.  For similar studies in the Sakai Sugimoto model see \cite{Bergman:2008sg,Bergman:2008qv,Kim:2008zn}.

In the electric case, for small enough ratios of the quark mass to the electric field strength, the probe $Dq$-brane becomes unstable, the signature of this instability being a region on the brane's worldvolume where the lagrangian becomes complex \cite{Karch:2007pd,Albash:2007bq}. The solution to this pathology is the introduction of a current in the direction of the electric field. Though this removes the pathology, the remnant of this is a submanifold of the $Dq$-brane worldvolume where the lagrangian density vanishes on-shell. This submanifold is known in the literature variously as the ``singular shell" \cite{Erdmenger:2007bn} and the ``ergosphere" \cite{Filev:2009xp} (due to its interpretation in the $T$-dualised geometry). It turns out that for any given electric field strength, there is a unique current which will cure the pathology. The ratio of the current to the electric field then provides a definition of the D.C. conductivity, as would be computed via a simple Ohm's law experiment.

Although the singular shell cures the problem of having an ill-defined action, it does introduce new complications. In particular in order to compute the spectral function for excitations on top of classical embeddings  which pass through the singular shell, the boundary conditions for the excitation, which are non-trivial, must be understood. In this paper we solve this problem, thereby allowing us to calculate more physical quantities in this interesting setup. 

A check that we have the correct boundary conditions will be to compare the D.C. conductivity calculated from the macroscopic setup \cite{Karch:2007pd} to that calculated using the Kubo relation, which involves the excitations on top of the classical brane embedding. Once the correct boundary conditions have been found we will be able to calculate the spectral function for  light-like momenta and thereby study the photoproduction rate for the plasma in the presence of an electric field.

The parameter space that we are interested in is spanned by the magnitude of the electric field, the quark mass and the charge density, and in the course of studying the spectral function we will discover a region of particular interest in it. In this region there is a very rapid change in behavior of the spectral function, going quickly from smooth oscillations to sharp spikes as we traverse this region of the tunable parameters. The change in behavior can be seen in several distinct physical quantities, including the conductivity, the susceptibility and the photoproduction rate. The behavior can be traced back to a rapid crossover in the embedding solutions which affects all subsequent quantities.

The outline of the paper is as follows: In section \ref{setup} we introduce our holographic setup. Wherever possible we work within the framework of the general $Dp/Dq$ flavor brane intersection. In this context we derive the constants of motion and the conductivity as given in \cite{Karch:2007pd} . 

In section \ref{specd3d7} we turn to the specific case of the $D3/D7$ intersection and illustrate the various types of brane behavior present by a series of plots. We then examine the change in behavior of the profiles upon a continuous variation of the electric field and the quark mass, as described above, and find the region where there is a quick crossover in behavior.
We also study the effect of this transition on the chiral condensate, charge susceptibility, conductivity and diffusion which can be derived directly from the background embedding solutions.

In section \ref{sec.fluc} we move on to the computation of correlators. We do this for transverse fluctuations only, propagating in the direction of the background electric field. We show that this is enough to compute the conductivity using the Kubo relation. This  can be done analytically for general $Dp/Dq$ systems and we find perfect agreement with the expression in \cite{Karch:2007pd}.

In section \ref{numerical}, in order to calculate the correlation functions away from the hydrodynamic limit we again specialize to the case of the $D3/D7$ setup. We study the transverse correlator at zero momentum for different values of the electric field and see that the crossover in the embedding behavior has a large effect on the spectral function. From the same function evaluated on the light-cone we obtain the photoproduction rate.  On integrating this magnitude to obtain the total luminosity we find a dramatic increase in the total number of emitted photons upon traversing the crossover region. 

We finally summarize our results and comment on possible extensions to the current work. 

\section{Holographic setup: background solutions}\label{setup}

In this section we summarize the framework used to study the embedding of $N_f$ probe $Dq$-branes in the background of a stack of $N_c \gg N_f$ coincident $Dp$-branes. This has been studied in many contexts in the literature, and a review of such investigations can be found in \cite{arXiv:0711.4467}. We will start here by detailing the generic $Dp/Dq$ intersection and later specialise to the $D3/D7$ system.

On a stack of coincident $Dq$-probe branes there is a natural $U(N_f)$ global symmetry whose abelian center can be identified with a baryonic $U(1)$ symmetry. The dynamics of the brane embeddings and of the $U(1)$ gauge field is dictated by the probe $Dq$-brane DBI action
\be\label{DBI}
\SS=-N_f T_{D_q} \int_{D_q} \dd x^{q+1} e^{-\phi} \sqrt{-\det(g_{ab} + 2 \pi \alpha' F_{ab})},
\ee
where $T_{D_q} = 1/((2\pi \sqrt{\alpha'})^q g_s \sqrt{\alpha'})$ is the $Dq$-brane tension, $g_s$ the string coupling, $\alpha'$ the inverse string tension and $g_{ab}$ the pullback metric from the 10-dimensional background. Here we will not deal with isospin degrees of freedom, hence any possible non-abelian effects are ignored. The background in which we wish to place these probe branes is the near-horizon limit of a stack of non-extremal $Dp$-branes
\beqalabel{background}
ds_{10}^2&=&G_{AB}\dd X^A \dd X^B  =  H^{-\frac{1}{2}} \left( -f(r) \dd x_0^2 + \dd \vec{x}_p^2 \right) + H^{\frac{1}{2}} \left( \frac{\dd r^2}{f(r)} + r^2 \dd \Omega_{8-p}^2 \right)\, ,  \,\,\,\,\,\,\,\\
e^\phi & = & H^{(3-p)/4}\, , \hspace{1cm} C_{01\cdots p}  =  H^{-1}\, , \\
H(r) & = & \left(\frac{R}{r} \right)^{7-p}\, , \hspace{1cm} f(r)  =  1-\left(\frac{r_h}{r} \right)^{7-p}\, ,
\eeqa
and the Hawking temperature is given by
\be
T=\frac{7-p}{4\pi R}\left( \frac{r_h}{R} \right)^{\frac{5-p}{2}}.
\ee

The probe branes wrap an n-sphere in the directions transverse to the $Dp$-branes, so we write
\be
\dd \Omega_{8-p} = \dd \theta^2 + \sin^2\theta \dd \Omega_n^2 + \cos^2\theta \dd \Omega_{7-p-n}^2\, ,
\ee
setting the classical $Dq$-brane embedding by a functional dependence\footnote{this is possible by exploiting an $O(8-p-n)$ symmetry.} $\cos\theta\equiv\psi(r)$. 
The  $Dq$-brane pullback metric is written as
\be\label{induced}
\dd s^2=g_{ab} \dd x^a \dd x^b = g_{00} (r) {\rm d}{x_0}^2 + g_{ii} (r) {\rm d} \vec{x}_p^2 + g_{rr} (r) {\rm d}r^2 + g_{\Omega\Omega} (r) {\rm d}\Omega_n^2\, ,
\ee
where the components of the induced metric $g_{ab}$ will coincide with the components of the background 10-dimensional metric $G_{ab}$ except for the radial term, given by
\be
g_{rr} = G_{rr} + G_{\theta\theta}(\psi) \psi'(r)^2 = \frac{H^{1/2}}{f(r)} \left( 1+ r^2 f(r) \frac{\psi'^2 }{1-\psi^2}  \right)\, .
\ee
All of the following analytic results will hold for any AdS black hole metric whose time component goes to zero linearly at the horizon, independent of the coordinate system chosen. All physical results will also be independent of redefinitions of the radial coordinate.
Here we will consider both finite charge density and a finite electric field directed along one of the spatial directions of the Minkowski space, say $x^p$. The holographic ansatz for the world-volume gauge field that takes these  two contributions into account is
\be
A =A_0(r) \dd x^0+ \left( E_p x^0 + A_p(r)\right) \dd x^p\, .
\ee
$A_0(r)$ and $A_p(r)$ will be normalizable solutions to the equations of motion, and their leading terms in an expansion around the asymptotic AdS boundary correspond  to the vacuum expectation values (${\it vev}$s) of current densities  $J^0(x)$ and $J^p(x)$ respectively.
The non-normalizable piece, $E_p x^0$, is dual to the source of the current, namely the electric field itself.

For compactness we will define a new matrix given by $\gamma_{ab}=g_{ab}+(2\pi \alpha')\, F_{ab}$. The only non-diagonal components of this matrix, $\gamma_{ij}$, are found when  $i,j$ take values in $i,j\in\{ 0, p, r \}$. We also define $\gamma^{ab}=\gamma^{-1}_{ab}$ and $\gamma\equiv \det\gamma_{ab}$.

The DBI action for our $Dp/Dq$ system is given by equation (\ref{DBI}), and can be expressed as
\footnote{we will consider only fluctuations which are independent of the Wess-Zumino term.} 
\be
\SS= -N_f T_{D_q}  \int \dd^{p+1} x\, \dd r\,\dd \Omega_{n} \, e^{-\phi} \sqrt{- g_{ii}^{p-1} g_{\Omega\Omega}^n \,W}\, ,
\ee
where the coefficients $g_{ab}$ are those of equation (\ref{induced}) and
\ben
W(r) = \det \gamma_{ij} = g_{rr} \left( g_{00} g_{ii} + {(2\pi \alpha')}^2 {E_p}^2 \right) +  {(2\pi \alpha')}^2 \left( g_{ii} A_0'^2 + g_{00} A_p'^2 \right)\, .
\een

There are two conserved quantities, $n_q$ and $j_p$, related to the charge density and the electric current respectively and defined by
\ben
\frac{\delta \SS}{\delta A_0'} = \frac{n_q}{\Omega_n}\, , \hspace{2cm} \frac{\delta \SS}{\delta A_p'} = \frac{j_p}{\Omega_n}\, ,
\een
where $\Omega_n$ is the volume of the unit $n$-sphere. The gauge field components can then be solved for as follows
\beqalabel{gaugesolution}
A_0' & = & - \frac{n_q}{\sqrt{g_{ii}} } \sqrt{-\frac{g_{rr} g_{00} \left( g_{00} g_{ii} + {E_p}^2 {(2\pi \alpha')}^2  \right)}{\NN^2 e ^{-2\phi}  g_{00} g_{ii}^{p} g_{\Omega\Omega}^n +  {(2\pi \alpha')}^2 \left( g_{00} n_q^2 + g_{ii} j_p^2\right)} }\, , \label{bgeqA0}  \\
A_p' & = &  \frac{j_p}{n_q} \frac{g_{ii}}{g_{00}} A_0'\, ,\label{bgeqAp}
\eeqa
where $\NN=N_f T_{D_q} \Omega_n  (2\pi \alpha')^2$.
As $g_{00}< 0$ (provided we are outside any possible horizons in the geometry), there is a radius $r_\star({E_p})$ such that $g_{00} g_{ii}(r_\star) + {E_p}^2 {(2\pi \alpha')}^2  = 0$.  This  characterizes a hypersurface of constant $r=r_\star$. At this point the legendre transformed action vanishes on-shell. We  will refer to it hereafter as the ``singular shell".  Calling $r_h$ the radius of the horizon, one can easily see that $r_\star({E_p})\geq  r_h$,  and  $ r_\star(0)= r_h$.
The Legendre transform of the DBI action is $\LSS  \sim - \int \dd^{p+1} x \dd r \tilde{\cal L}_{DBI}$ with
\be\label{legtrans}
\tilde{\cal L}_{DBI} =-e^{-\phi}\sqrt{-(g_{00}g_{ii}+ {(2\pi \alpha')}^2 E_p^2)g_{ii}^{p-1}g_{rr}g_{\Omega\Omega}^n} \sqrt{1+\frac{ (2\pi \alpha')^2( g_{00} n_q^2 +  g_{ii} j_p^2 )}{\NN^2 e^{-2\phi}g_{00}g_{ii}^p g_{\Omega\Omega}^n} }\, .
\ee
Furthermore,  as the quantity $g_{00} g_{ii} + {E_p}^2(2\pi \alpha')^2$ changes sign at $r=r_\star$, in order for (\ref{legtrans}) to remain real, we impose that the second square root in it also changes sign at $r_\star$. This leads to a relation\footnote{it is interesting to note that in 
\cite{Herzog:2006gh,Gubser:2006nz,CasalderreySolana:2007qw}, a remarkably similar structure to the singular shell was observed in the trailing string solution, related to the local speed of light. Clearly the situations of a single moving quark and the quark current studied here are extremely similar. One can show that the infinite mass limit of the present current coincides with the current of a single quark moving through the plasma. We would like to thank C. Herzog and J. Casalderrey for pointing out this interesting observation.} between $j_p$ and $E_p$:
\be
j_p =  {E_p} \sqrt{ \NN^2  e^{-2\phi}   g_{ii}^{p-2 } g_{\Omega\Omega}^n + n_q^2 (2\pi \alpha')^2 g_{ii}^{-2} }\Bigg|_{r_\star}, \label{jpofEp}
\ee
and defining  the conductivity as usual through Ohm's law, we get $\sigma=j_p/{E_p}$. This is the conductivity first found in \cite{Karch:2007pd} and the result is calculated purely in terms of background quantities. Although compact, the notation should not hide the important fact that $\sigma$ is a non-linear function of $E_p$ (see (\ref{current}) below). This dependence arises from the evaluation of induced metric components at the singular shell $r_\star$. By studying linearized fluctuations on top of these background embeddings we will be able to extract information about properties of the gauge theory, and also calculate the conductivity using the Kubo formula. Agreement with the conductivity found in the macroscopic Ohm's law calculation (eq. \ref{jpofEp}) will be a good check of our methods (see section \ref{sec.fluc} for details).

\section{Specializing to the $D3/D7$ system \label{specd3d7}}

In order to picture what is happening in the general $Dp/Dq$ intersection we discuss here the possible embeddings in the case of the $D3/D7$ solution. In future sections we will discuss fluctuations about these solutions so we now introduce the coordinate system which will be made use of throughout. We find it convenient at this stage to use the radial variable, $u$, defined as $u= (r_h/r)^2$.
The pullback metric of the $D7$-brane onto the $D3$-brane background will then be given by
\begin{equation}
\frac{ds^2}{R^2}=\frac{\pi^2T^2}{u}\left((u^2-1) \dd t^2+\dd \vec x_{3}^2\right)+\frac{(1-\psi^2+4u^2(1-u^2)\psi'^2)}{4u^2(1-u^2)(1-\psi^2)} \dd u^2+(1-\psi^2) \dd\Omega_3^2\, ,
\end{equation}
where $u=u_b=0$ corresponds to the UV of the gauge theory (the boundary of AdS) and $u=u_h=1$ is the black hole horizon. The background field $\psi(u)$ corresponds to the classical embedding of the $D7$-brane, and the gauge field, including transverse perturbations, is given by
\be\label{backgroundgauge}
A=  A_0(u) \dd t + \left( E_z \,t+A_z(u) \right)\dd z+ \dA_\perp(t,z,u) \dd x_\perp\, ,
\ee
with $x_\perp=\{Êx,yÊ\}$. Here $A_0(u)$ corresponds to the non-zero baryon density, $A_z(u)$  to the non-zero current in the $z$-direction, and $E_z$ is the finite electric field in the $z$ direction.  In this paper we will deal only with the transverse fluctuations, captured entirely in $\dA_\perp(t,z,u)$ and defer the full investigation of the longitudinal and scalar modes for future study. The transverse sector decouples completely as we consider momentum in the direction parallel to the background electric field (see section \ref{sec.fluc} for further details).
It is convenient to define the dimensionless electric field $\efield$ as
\be
\efield \equiv \frac{ {(2\pi \alpha')} R^2}{\left(\pi T R^2\right)^2} E_z\, ,
\ee
and constants of motion $\bardens$ and $\current$, related to the charge density and current  given in the previous section by
\be
\begin{pmatrix} \bardens \\ \current \end{pmatrix} \equiv 
\frac{4  {(2\pi \alpha')}}{N_c N_f T^2\left( \pi T R^2 \right)}
\begin{pmatrix} n_q \\  j_z \end{pmatrix} \, .
\ee
However there is only one value of the current which renders a well defined action inside the singular shell. The position of the singular shell in this case is given by $u_\star=1/\sqrt{1+{\efield}^2}$, and the current is found by demanding reality of the action across the singular shell, leading to
\be\label{current}
\current={\efield}\sqrt{\sqrt{1+{\efield}^2}(1-\psi_\star^2)^3 + \frac{\bardens^2}{1+{\efield}^2}}\, ,
\ee
where $\psi_\star=\psi(u_\star)$. From this we can read off the electrical conductivity
\be
\label{dcconduct}
\sigma=\frac{j_z}{E_z} = \frac{N_c N_f T}{4Ê\pi} \frac{\current}{\efield}  \equiv \frac{N_c N_f T}{4 \pi} \dlcond\, .
\ee

Using equation (\ref{current}) we can express the background gauge fields $(A_0(u), A_z(u))$ entirely in terms of $\bardens,\efield$ and the probe brane embedding profile $\psi(u)$
\begin{eqnarray}\label{eq.Ap}
A_0'(u)^2 &=& \frac{(\pi T R^2)^2 \bardens^2 \left(1-u^2 \left(1+{\efield}^2\right)\right)  \left(1-\psi^2+4 u^2 \left(1-u^2\right) {\psi '}^2\right)}{4(2\pi \alpha')^2 \left(1-\psi
  ^2\right)  \left( (1-u^2) \left( 1-\psi^2 \right)^3 - u^3 \left(\current^2-\frac{2\bardens^2}{1+{\efield}^2}\right) -  u^5 \bardens^2  \right)} \, , \nonumber \\
A_z'(u)^2 &=& \frac{\current^2}{\bardens^2} \frac{A_0'^2}{(1-u^2)^2} \, .
\end{eqnarray}
From here, $A_0(u)$ can be obtained by integrating from the horizon $u=1$ with boundary condition $A_0(1) =0$. The UV series solutions to these gauge fields are given by
\beqa\label{defmu}
A_0&=&\mu-\frac{TR^2}{4\alpha'}\bardens u+...\, ,\\
A_z&=&\frac{TR^2}{4\alpha'}{\tilde{j}}_z u+...\, ,
\eeqa
$\mu$ being the chemical potential canonically conjugate to the baryon density.

The background field solution $\psi(u)$ corresponds to the classical embedding of the $D7$-brane. The range of this field is $0\le\psi(u)\le1$ and it is completely determined by its value either on the singular shell, or, for ``Minkowski embeddings", the value of $u$ for which this field goes to $1$. The asymptotic behavior of the embedding near the UV gives the usual source (mass) and ${\it vev}$ for the quark bilinear operator, leading to the relation that
\begin{equation}
\psi(u)_{u\rightarrow 0}\simeq \frac{m_q}{\sqrt{2}} u^\frac{1}{2} + \frac{c_q}{2\sqrt{2}}  u^\frac{3}{2}\, .
\end{equation}
The quark mass and condensate being given by
\beqalabel{masscond}
M_q & = & \frac{1}{2} \sqrt{g_{YM}^2 N_c} T m_q\, , \\
\left< \bar \Psi \Psi \right> & = & - \frac{1}{8} \sqrt{g_{YM}^2 N_c} N_f N_c T^3 c_q\, .
\eeqa
Although this is really the source and ${\it vev}$ of a supersymmetric operator \cite{hep-th/0701132}, in the present case where supersymmetry is broken, both by the electric field and by the finite temperature, we will simply refer to the operator in question as the quark bilinear.

In what follows we shall plot the classical solutions to the embedding equation in a cartesian coordinate system in which it is more intuitive to picture the flows of the $D7$-brane. The change of variables between the original coordinates and those we will use to plot the solutions is
\ben
(\rho,L)=\left( \sqrt{1-\psi (u)^2},\psi(u) \right)\frac{\sqrt{\sqrt{1-u^2}+1}}{\sqrt{u}},
\een
Notice that $\psi = \cos\theta$ and the global factor defines an isotropic  radial coordinate, $w(u) = \sqrt{\rho^2 + L^2}$, as used in many papers
(see for example \cite{arXiv:0711.4467}).

\begin{figure}[ht]
\begin{center}
\includegraphics[scale=0.75]{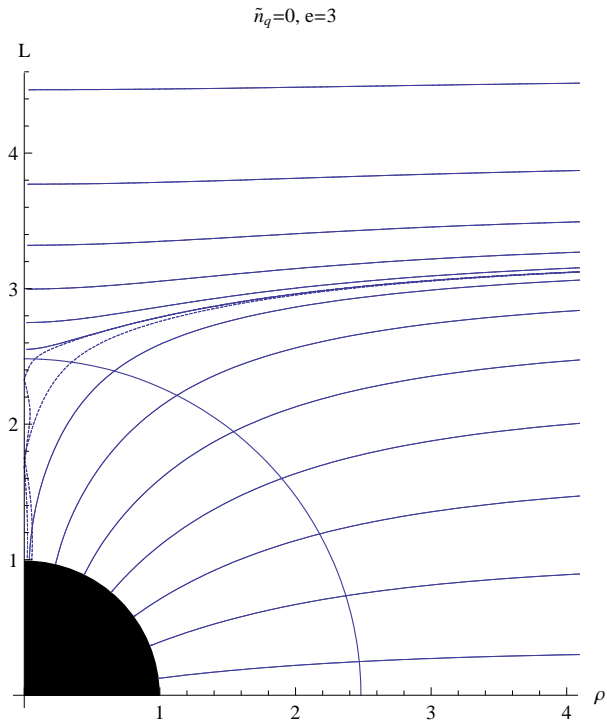}
\caption{\em \label{zerodisemb}
Embeddings at zero baryon density showing the Minkowski embeddings, which do not hit the singular shell (in this case at $\rho^2+L^2=3+\sqrt{10}$) conical embeddings given by short dashed lines which pass through the singular shell and hit the $L$-axis, and black hole solutions which pass smoothly from the singular shell to the horizon.
}
\end{center}
\end{figure}

For the $D7$ embedding in the finite temperature $D3$ geometry with finite electric field there are three families of solutions illustrated in figure \ref{zerodisemb}. We discuss here the three types of behavior labelled ``Minkowski solutions", ``Black hole solutions" and ``Conical solutions".

{\em \textbf{Minkowski solutions:}} These solutions  do not intersect the singular shell and therefore do not meet the horizon. The current in this case is zero as there are no free charge carriers. For these solutions the energy supplied by the electric field is not sufficient to pair create quarks of such a high mass. They therefore only appear for sufficiently high values of $m_q/T$. Fundamental matter remains bound in long-lived mesons.

Minkowski solutions are labelled by the point of closest radial approach to the $D3$ branes, $u=u_{max}$, at which they reach $\theta=0$ or $\psi(u_{max})=1$ with  $u_h >u_\star \ge u_{max}$. The second boundary condition needed to fix these solutions determines that there is no conical singularity at $u_{max}$, corresponding to  $\psi'(u_{max})=-\infty$.

\begin{figure}[ht]
\begin{center}
\includegraphics[scale=0.7]{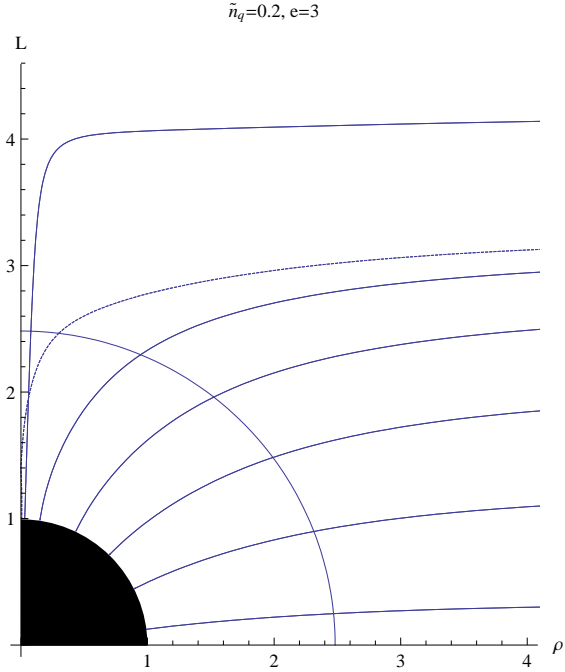}
\caption{\em \label{finitedisemb}
Embeddings at finite baryon density and the same electric field as in the previous figure illustrating the absence of Minkowski solutions. Note that there are still conical solutions in this case. 
}
\end{center}
\end{figure}

{\em \textbf{Black hole solutions:}} These solutions pass smoothly through the singular shell and end on the black hole horizon without ever reaching $\psi(u)=1$ for any value of $u$. The fact that the solutions pierce the singular shell indicates that there is a finite current. For these solutions, the quark mass is sufficiently small that the electric field can pair-create them. With these free charges  a current can be set up. With an external electric field there is an unlimited supply of energy and one might have thought that the charge carriers would continue to accelerate. However, in the presence of the adjoint matter (with $N_f/N_c\rightarrow 0$) the energy of the fundamental matter is lost into this background and an equilibrium situation with a fixed, finite current is set up \cite{Karch:2007pd}. At finite baryon density all solutions have to end on the horizon (figure \ref{finitedisemb}).

\begin{figure}[ht]
\begin{center}
\includegraphics[scale=1.4]{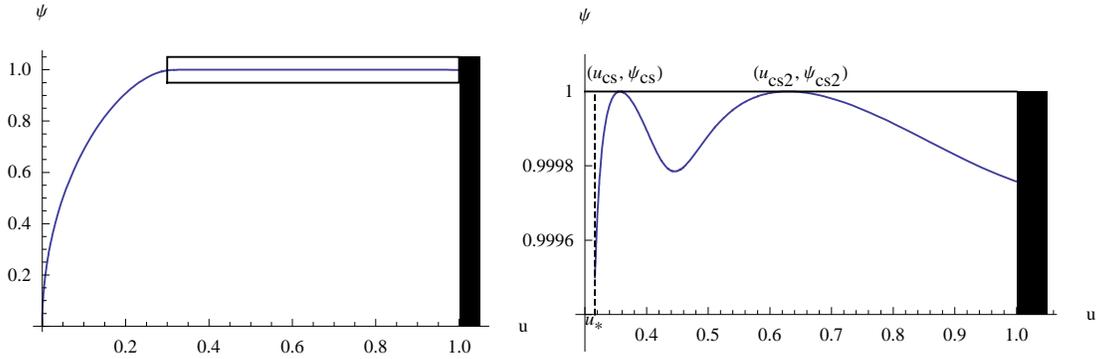}
\caption{\em \label{conicalexample}
Example of a solution with two conical singularities. The left hand plot includes the whole range of the embedding from the UV ($u=0$) to the horizon ($u=1$). The second part focuses just on the region around the conical singularity, illustrated in the left hand plot by the boxed region. Here the first conical singularity is labeled as $(u_{cs},\psi_{cs})$ while the second is labeled as $(u_{cs2},\psi_{cs2})$. The $L$-axis is the line $\psi=1$ while the horizon is marked as the black box. The singular shell is marked in the second plot as $u_\star$. 
}
\end{center}
\end{figure}

{\em \textbf{Conical solutions:}} These pass through the singular shell and appear to display a conical singularity at a radius, $u_{cs}<u_h$, determined by $\psi(u_{cs})=1$. We illustrate one of these solutions in figure \ref{conicalexample} in the original ${u,\psi}$ variables.

Solutions which pass through the singular shell are determined uniquely by specifying $\psi_\star=\psi(u_\star)$. All derivatives are fixed by this single boundary condition and the singular shell acts as an attractor for solutions\footnote{there are four solutions to the expansion about the singular shell, two of which are complex and one of which gives non-physical, oscillatory behavior.}. This is reminiscent of the black hole horizon itself in the zero electric field situation, with which one can fix the behavior of a solution in the UV uniquely with a single boundary value. 

The electric field lines caused by the presence of free charges need to end  somewhere and, hence, the brane must  intersect the horizon. 
For the conical embeddings, the same reasoning applies. 
Solving numerically we find that these solutions continue past the point of the conical singularity until they reach the horizon. Still, in $(\rho,L)$ coordinates, at the points where they touch the $L$-axis one can observe that there is a jump in  slope, hence the name conical singularity. After reaching this point, the value of $\psi$ starts to decrease again, and the brane eventually hits the horizon (possible after encountering another conical singularity as illustrated in figure \ref{conicalexample}). By studying the tension of the $D7$-brane at the conical singularity we find that there is no other object which can be attached to the $D7$-brane matching it, meaning that the physical configuration is the one where the $D7$-branes continues through the conical singularity, eventually reaching the horizon.

\subsection{Singular shell solutions}
The solutions which pass through the singular shell exhibit subtle behavior which will influence much of the physics in what is to come. For this reason we will now discuss the details of the singular shell embeddings, parametrised by the three variables $m_q$, $e$ and $\bardens$. 

Especially interesting is the behavior in the $(m_q, \efield)$ plane. First we chose to fix the electric field and look at the solutions as a function of $m_q$.

\begin{figure}[ht]
\begin{center}
\includegraphics[scale=1.4]{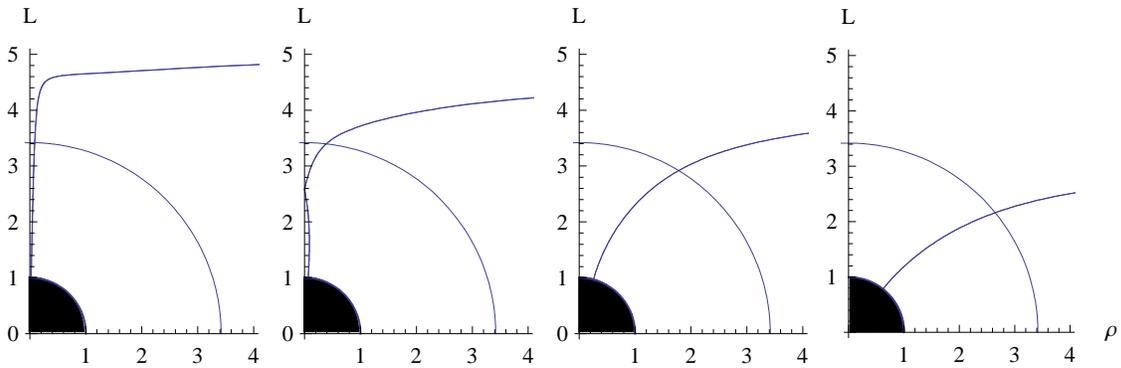}
\caption{\em \label{embvmq}
Evolution of the embedding profiles with decreasing values of the  mass $m_q=5,\, 4.5,\, 4$ and $3$ at fixed values of $\bardens=0.05$ and electric field  $\efield=5.8$. After developing conical singularities the profile spreads suddenly for lower masses.}
\end{center}
\end{figure}

In figure \ref{embvmq} we see the behavior for a range of masses. 
For the largest mass, $m_q=5$, we see that there is a narrow throat, which leads from a relatively straight $D7$-embedding, sharply down into  the singular shell and subsequently the horizon. As we decrease the mass to $m_q=4.5$ we see that the behavior changes and we obtain a solution with a conical singularity. At this point the brane has a smoother profile and the sharp throat has disappeared. For still smaller masses than $m_q=4$ we return to the pure black hole embedding and the conical singularity disappears. The profile is much smoother and there is a ``wide throat" to the solution, rather than the previous sharp spike. Decreasing $m_q$ still further widens this throat slowly and monotonically. 

It should be noted that, with decreasing $m_q$, the conical solutions start with the conical singularity at the point where the black hole horizon meets the $L$-axis ($u=u_h$). The position $u_{cs}$ of the first conical singularity (for there may be multiple) moves away from the horizon and then returns to it as the quark mass is decreased further.

We see from this example that over a small range of masses there is a very fast change in behavior, from a sharp spike solution, through a conical solution, to a smooth, wide throat solution. This behavior is generic, though the region over which this fast transition occurs is dependent on both $m_q$ and $\efield$. Indeed not all values of $\bardens$ and $e$ will exhibit conical solutions.

\begin{figure}[ht]
\begin{center}
\includegraphics[scale=1.4]{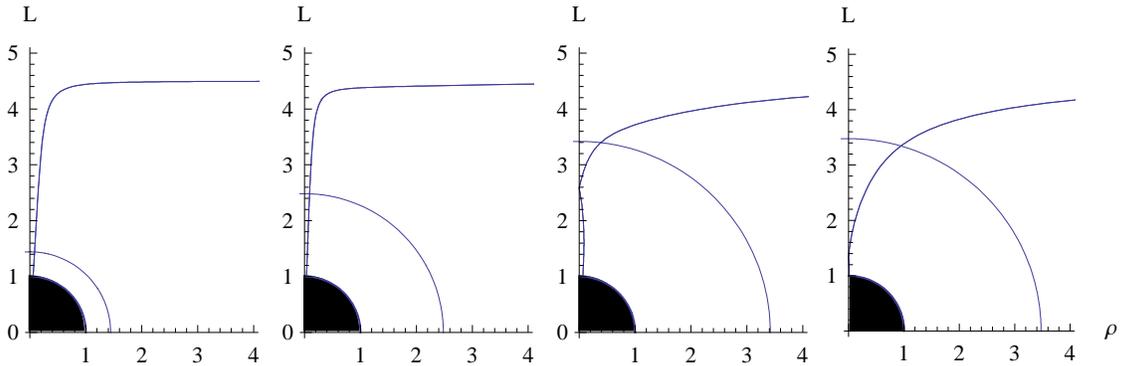}
\caption{\em \label{embvee}
Evolution of the embedding profiles with the electric fields $\efield= 0.8, 3, 5.8$ and $6$ at fixed values of $\bardens=0.05$ and quark mass $m_q=4.5$. We observe the developement of conical embeddings in the range $\efield \in (5.7,6)$. Increasing the value of $e$  there is a sudden change in form, and the profiles spread significantly.}
\end{center}
\end{figure}

Keeping $m_q$ fixed and varying $e$ gives a similar behavior, exhibited in figure \ref{embvee}. We label the region where the behavior changes quickly as a function of $m_q$ or $\efield$ the ``crossover region". As stated above, this region may or may not contain a subregion of conical singular solutions, depending on the value of $\bardens$.

As a physical motivation for such a crossover it seems that there are two competing mechanisms forcing the brane embedding to go in two different directions. The crossover is the point where one of of these effects becomes dominant over the other and there is  a sudden change in the behavior. This is rather reminiscent of the two effects seen in the expression for the conductivity, one of which becomes more important as $\efield$ increases and one of which becomes more important as it decreases. These two terms are related to the the pair-creation and presence of free charges respectively. 

\begin{figure}[ht]
\begin{center}
\includegraphics[scale=0.65]{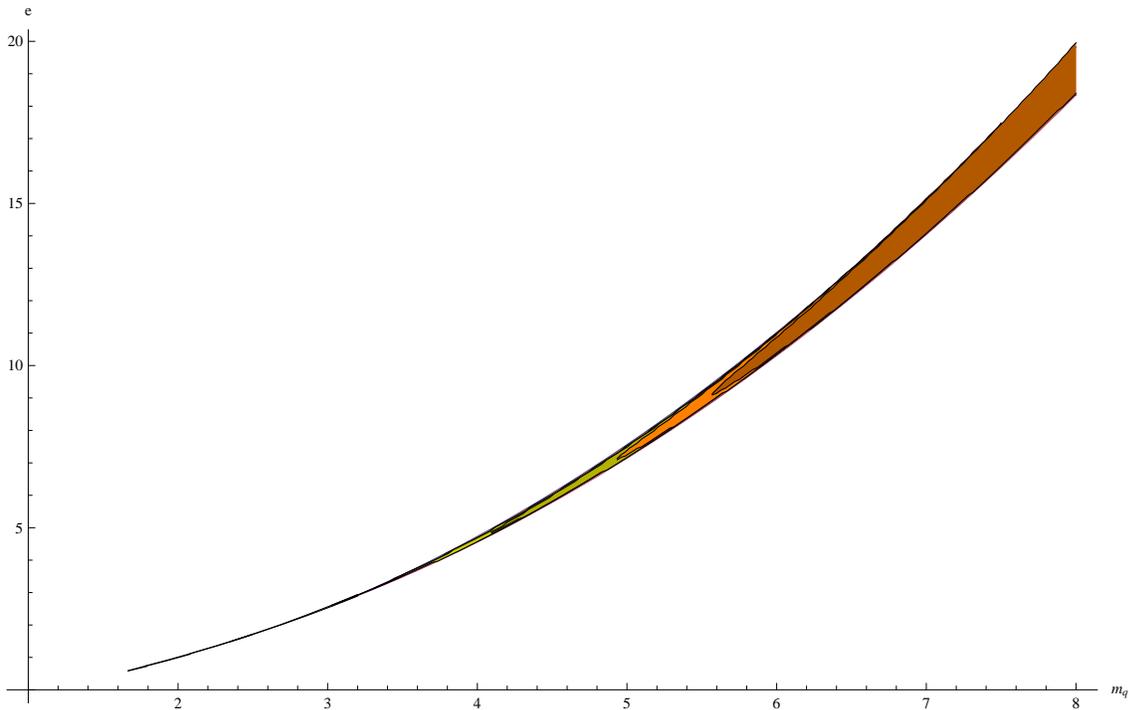}
\caption{\em \label{transregion}
The shaded region is where conical embeddings are found in the plane $(m_q, \efield)$. The regions for different $\bardens$ are superimposed, beginning at larger $m_q$ the larger $\bardens$ is. The lowest region is for $\bardens=2\cdot 10^{-5}$ and this contains all other regions for larger values of $n_q$. The orange region is for $\bardens=0.6$ and the brown one corresponds to $\bardens=1.2$.}
\end{center}
\end{figure}

We can plot the crossover region in the $(m_q,\efield)$ plane in which conical solutions are present.  This wedge is shown in figure \ref{transregion} for varying values of $\bardens$ corresponding to the increasingly dark colors. It is rather peculiar that the only effect of the baryon density is to shift the starting point within the same region and not to displace the region at all in some direction.

\subsection{Chiral condensate}
Having detected a crossover, given by a sudden change in the behavior of probe embeddings, it is natural to expect that other physical quantities will be affected. In the present section we will examine the chiral condensate.

\begin{figure}
\begin{center}
\includegraphics[scale=0.7]{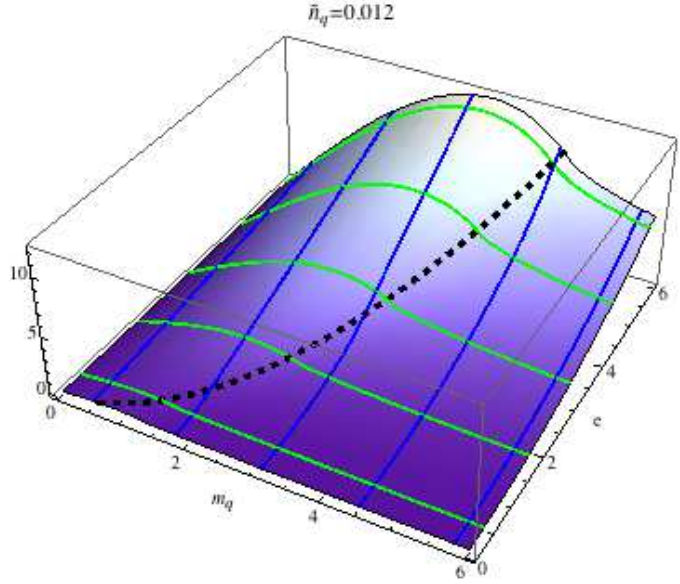}
\caption{\em \label{condenstrans}
Condensate as a function of electric field and mass for $\bardens=0.012$. Curves of constant $m_q$ are plotted in blue, and those of constant $\efield$ in green. We superimpose in black the curve where conical embeddings occur. Upon crossing it there is a mild but noticeable change in slope.
}
\end{center}
\end{figure}

We plot $-c\sim\langle\bar\Psi\Psi\rangle$ given by equation (\ref{masscond}) as a function of $m_q$ and $\efield$ for $\bardens=0.012$ in figure \ref{condenstrans}. We see that in the region where the crossover in behavior occurs at the singular shell the condensate also undergoes a faster change in behavior, as compared with other regions in the $(m_q,\efield)$ plane. We have superimposed on the graph the curve $\efield =0.29 \,m_q^2$ which runs through the middle of the orange region in figure \ref{transregion}. We see that this curve coincides roughly, as expected, with the region where the level curves change from concave to convex. We should note however that the change in behavior of the chiral condensate over the crossover region is much less pronounced than that of many other quantities as we will show. 

\subsection{Charge susceptibility}\label{sec.susc}

\begin{figure}[ht]
\begin{center}
\includegraphics[scale=0.7]{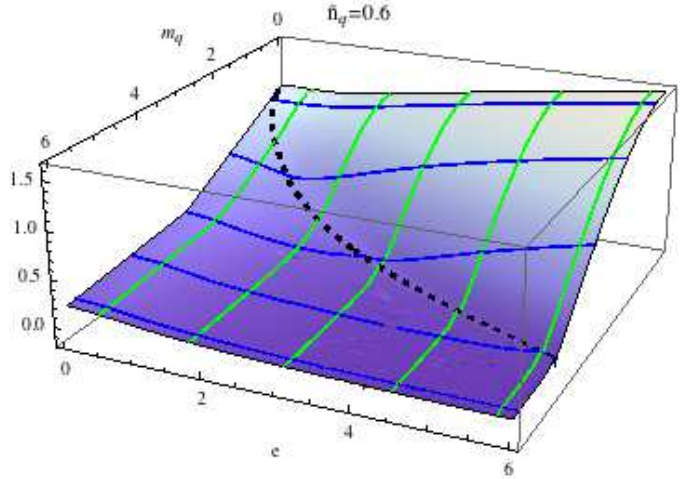}
\caption{
\em \label{susceptibility}
Dimensionles charge susceptibility $\tilde \Xi$ as a function of the
quark mass and electric field for $\bardens=0.6$. Green, blue and
black lines have the same meaning as in the figure for the chiral condensate.
}
\end{center}
\end{figure}

Another physically relevant quantity showing a fast change in behavior
across the crossover described in the previous sections is the
charge susceptibility, thermodynamically defined as
\be\label{eq.susc1}
\Xi = \frac{\partial n_q}{\partial \mu} \Bigg|_{T}  = \frac{N_c N_f
T^2}{2} \frac{\partial \bardens}{\partial \tilde \mu} \Bigg|_{T} \equiv
\frac{N_c N_f T^2}{2} \tilde \Xi,
\ee
where $\mu$\ is defined\footnote{where we have defined $A_0=\frac{TR^2}{4\alpha'}(\tilde\mu-\bardens u+\cdots)$ from equation (\ref{defmu}).} in equation (\ref{defmu}). From eq. (\ref{bgeqA0})
we can read the dependence of $\mu$ on the charge density at
fixed temperature and extract the value of the susceptibility. In
figure \ref{susceptibility} we plot this quantity as a function of the
quark mass and the electric field for $\bardens=0.6$. There we can see
that in the region where we have narrow throats in the embedding
profiles the susceptibility varies very slowly. On the other hand, in
the region with wide-throat profiles we have a more pronounced slope.
These two regions are separated by the line where conical embeddings
may occur. We will find this behavior again, when considering the
conductivity and the photoproduction.

\subsection{Conductivity}

\begin{figure}[ht]
\begin{center}
\includegraphics[scale=0.55]{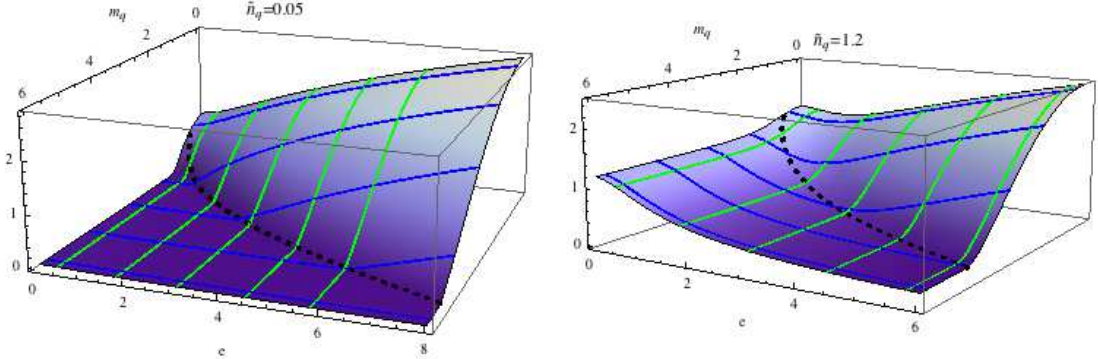}
\caption{\em \label{conductplot}
Dimensionless conductivity, $\tilde \sigma$, as a function of $\efield$ and $m_q$ for two different values of $\bardens = 0.05$ and $1.2$. Varying $m_q$ at constant $\efield$ shows a stronger change in  slope upon crossing the black line than varying $\efield$.
 }
\end{center}
\end{figure}

In  \cite{Karch:2007pd}, by analyzing eq. (\ref{dcconduct}) it was conjectured that the two terms present under the square root have physically clear and distinct origins. The second one is the clearest, as it is directly related to the presence of charge carriers proportional to the net baryon charge $\bardens$.  Notice the presence of the electric field in the denominator, which suppresses this contribution  for larger values of $\efield$.
 The first one should correspond to the presence of charge carriers that come from pair production from the vacuum. 
It is enhanced by a stronger electric field, and, assuming a direct relation between $\psi_*$ and the mass, this
term should vanish for high enough masses.
Roughly speaking, this behavior is seen in the set of graphs shown in figure \ref{conductplot}. 

What is less obvious is the sharpness of the transition to the region of high conductivity dominated by the first term in (\ref{eqcond}). This is
the slope which is visible for high values of $\efield$ and low values of $m_q$. Clearly the abruptness in this transition comes from the hidden dependence of $\psi_\star$ on $\efield$ and $m_q$, and the sudden change in profiles that was investigated in the previous section is clearly behind this strong sensitivity. 

\subsection{Charge diffusion}

A formal study of the charge diffusion for this system would involve
solving for the longitudinal vector channel, that presents a pole in its
correlator which, in the hydrodynamic limit, gives information about
this transport coefficient. In this paper we do not follow this
procedure but exploit the Einstein relation, which expresses the
diffusion as the quotient of the conductivity by the susceptibility
\be
D= \frac{\sigma}{\Xi} = \frac{1}{2\pi T} \frac{\tilde \sigma}{\tilde
\Xi} \equiv \frac{\tilde D}{2\pi T}\, .
\ee

\begin{figure}[ht]
\begin{center}
\includegraphics[scale=0.7]{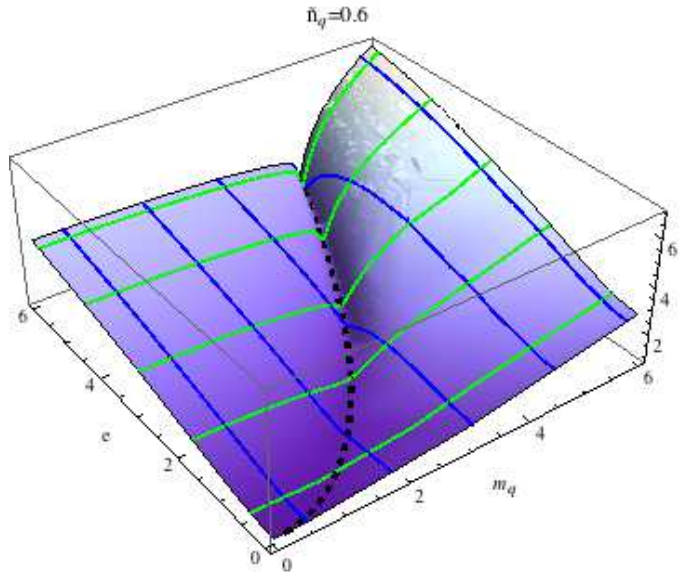}
\caption{
\em \label{diffusion}
Dimensionless charge diffusion $\tilde D$ as a function of the quark
mass and electric field for $\bardens=0.6$.  }
\end{center}
\end{figure}

With our numerical methods we are able to recover the vanishing electric field
results of \cite{arXiv:0811.1750}, and in figure \ref{diffusion} we show how these get
modified in the presence of a finite electric field for
$\bardens=0.6$.

In this figure we can see an interesting behavior. Keeping the
electric field fixed and studying the diffusion constant when the quark
mass increases we first see that it decreases very slowly until one
reaches the crossover region. Once this happens there is first a rapid
decrease and, once we pass this region and enter the parameter space
where we have narrow-throat embedding profiles, we find that the
diffusion constant increases with the quark mass very fast. This rapidly changing behavior with the mass is correlated with the vanishing of the susceptilibity in the same limit.

\section{Fluctuations}\label{sec.fluc}

In order to compute current-current correlators we must consider perturbations of the gauge field $\dA_\mu$. The components $\dA_0$ and $\dA_p$  are scalars under the little group $O(p-1)$, and will mix with any other scalar modes present in the system, for example the profile fluctuation 
$\delta\psi$ or fluctuations of the dilaton $\delta\phi$. This channel is fairly complicated to analyse and for the D3/D7 case it has been studied in the absence of electric field in \cite{arXiv:0805.2601}. On the other hand transverse fluctuations $\dA_\perp = \dA_i$ with $i=1,\cdots,p-1$ will transform as vectors, and hence decouple completely from all other excitation modes and have been studied, again in the absence of an electric field, in \cite{arXiv:0710.0334}. In this paper we will be concerned exclusively with these transverse fluctuations, leaving the coupled sector for future study.

In a rotationally invariant theory, the retarded Green's function $G^R_{\mu\nu}$ involving perturbations of the form   $\exp(-ik_\mu x^\mu)$ is generically split into two different polarization tensors \cite{hep-th/0607237}
\be
G^R_{\mu\nu} (k)=P^\perp_{\mu\nu}(k)\Pi^\perp(k) + P^{||}_{\mu\nu}(k)\Pi^{||}(k)\, ,
\ee
where the projectors are defined as $P_{00}^\perp=P_{0m}^\perp=0$, $P_{mn}^\perp=\delta_{mn}-k_m k_n/\vec{k}^2$, $P^{||}_{\mu\nu}=\eta_{\mu\nu}-k_\mu k_\nu/k^2$. In the present setup we expect this structure to remain unaltered as long as the background electric field, $E_p$, points in the same direction, $x^p$, as the propagation of the perturbation and, hence leaves the $O(p-1)$ invariance untouched.  For the transverse fluctuations, the components of $G^R_{rs}$ for $r,s = \{1,\cdots,p-1\}$
are related directly to $\Pi^\perp$ by
\be
G^R_{rs} (k)=\delta_{rs} \Pi^\perp(k)\, .
\ee
The prescription for the Lorentzian correlator relates the polarization function $\Pi^\perp(k)$ to the boundary action for the transverse fluctuations
\be\label{bouact}
S_{bou} = -\frac{\NN}{2} \int_{r\to r_b} \dd^{p+1}x\,  e^{-\phi} \sqrt{-\gamma} \gamma^{\pperp} \gamma^{rr}\dA'_\perp \dA_\perp\, ,
\ee
with $\NN$ defined as in equation (\ref{gaugesolution}). This prescription gives \cite{hep-th/0205051}
\be\label{twopoint}
\Pi^\perp = \NN e^{-\phi} \sqrt{-\gamma} \gamma^{\pperp} \gamma^{rr} \frac{\dA'_\perp \dA_\perp^*}{|\dA|^2}\Bigg|_{r\to r_b}\, .
\ee
In section \ref{numerical} we discuss further  how to calculate this quantity for the specific case of the $D3/D7$ system.

\subsection{Transport phenomena}

The conductivity  is defined macroscopically via Ohm's law as the ratio between the current density and the field that creates that flow of current. The calculation of the conductivity for a $Dp/Dq$ intersection has been done following this definition in \cite{Karch:2007pd}.  Here we are interested in a microscopic derivation using the Kubo relation:
\be\label{kubo}
D\Xi =  -\lim_{\omega\to0} \frac{{\rm Im}\Pi^\perp(\omega=q)}{\omega}\, ,
\ee
where $D$ is the diffusion constant and $\Xi$ the susceptibility of the medium. 
In the next section we will show that this prescription leads to the same analytical expression obtained in \cite{Karch:2007pd}, thereby proving the validity of the Einstein's relation  $\sigma=D\Xi$. 
An independent check of this relation amounts to calculating the susceptibility $\Xi$ and the diffusion constant $D$ separately.

The susceptibility is an equilibrium quantity given by a thermodynamic definition
$\Xi = \d n_q/\d \mu\Big|_T $,
where $\mu$ is the chemical potential and $n_q$ the charge density, as has been discussed in section \ref{sec.susc}. On the other hand, the diffusion constant, $D$, can be obtained from a hydrodynamic pole in the correlator for the longitudinal channel of vector fluctuations, relating the frequency and the momentum as dictated by Fick's law:
$
\omega = - i D q^2 + \OO(q^3)\, ,
$
although we leave this calculation for the future. Numerically one can see that in the case of vanishing electric field the product $ D\Xi$ obtained in this way agrees with $\sigma$, as explicitely shown in \cite{arXiv:0811.1750}.

\subsection{Gauge field fluctuations}

At this point we must study what boundary conditions we need to impose on the transverse fluctuations at any singular points in their equations of motion. The equation of motion for the transverse fluctuations after inserting the plane wave ansatz $\dA_\perp(t,\vec{x},r)\rightarrow e^{-i(\omega x^0-qx^p)}\dA_\perp(r)$ reads
\beqad
&&\partial_r \left[  e^{-\phi} \sqrt{-\gamma} \gamma^{\pperp} \left( \gamma^{rr} \dA_\perp' - i \left( \omega \gamma^{(0r)} - q \gamma^{(pr)} \right) \dA_\perp\right) \right] - i e^{-\phi} \sqrt{-\gamma} \gamma^{\pperp} \left(  \omega \gamma^{(0r)} - q \gamma^{(pr)} \right) \dA_\perp'  \nonumber \\  
&&- e^{-\phi} \sqrt{-\gamma} \gamma^{\pperp} \left( \omega^2\gamma^{00} +  q^2 \gamma^{pp} \right) \dA_\perp  + 2\, \omega\, q\, e^{-\phi} \sqrt{-\gamma} \gamma^{\pperp} \gamma^{(0p)} \dA_\perp =   0\, .\label{finiteEfluc}
\eeqad

At finite temperature we need to make sure that the solution is regular at the horizon\footnote{provided a non vanishing $n_q$ the branes can be black hole-type or conical-type. They touch the black hole horizon at some point, and the formula obtained will be shown to coincide with numerics also for conical-type embeddings.}. This can be achieved by performing a Frobenius expansion. The solution is regularised as $\dA_{_\perp}(r) = (r-r_h)^{\eta^{(h)}} \dA_{\perp,reg}$ where the indices are found to be
\be\label{firstindex}
\eta^{(h)}_{\pm} =  \frac{1\pm1}{2} - i\, \frac{\omega}{4 \pi T}\, ,
\ee
the $(h)$ indicating that this is an index at the horizon. Note that this is markedly different from the case of zero electric field where the only possibilities were outgoing or incoming boundary conditions. Now there is the possibility of an additional damping factor at the horizon. In this case the choice of outgoing or incoming boundary conditions is dependent on the relative sign between $A_0'$ and $A_p'$. In our choice of the background gauge fields (equation (\ref{gaugesolution}))  we have chosen the relative sign of these to give the appropriate incoming wave condition.

In addition to the singular behavior at the horizon, there is also a singularity in the equations of motion for the fluctuations at the singular shell, $r_\star$. We can perform a similar Frobenius study in this region. In this case we find one trivial index, $\eta^{(\star)}_0=0$, while the other, $\eta^{(\star)}_r$, is a very complicated function of $E_p$, $\omega$, $n_q$ and $\psi_\star$ which, in the limit of vanishing electric field, is given by $\eta^{(\star)}_{r}(E_p\rightarrow 0)=i\frac{\omega}{2\pi T}$.

Finally, in order to choose our boundary conditions we want to recover the usual expression for the indices at the horizon in the limit of ${E_p}\to 0$. Notice that in this limit, as we have already seen, $r_\star \to r_h$, so
\be
\left( r - r_h \right)^{\eta^{(\mathrm{lim})}} \equiv  \lim_{{E_p}\to 0} \left( r- r_h \right)^{\eta^{(h)}} \left( r- r_\star \right)^{\eta^{(\star)}} = \lim_{{E_p}\to 0} \left( r- r_h \right)^{\eta^{(h)}+\eta^{(\star)}} \overset{?}{=} (r-r_h)^{\pm i\frac{\omega}{4\pi T}}\, .
\ee
We have four possible combinations of the two values at the horizon and at the singular shell, combining to give the following values for $\eta^{(\mathrm{lim})}$.
\begin{table}[h]
\begin{center}
\begin{tabular}{l|cc}
$\eta^{(lim)}$ & $\eta^{(h)}_+$ & $\eta^{(h)}_-$ \\
\hline
$\eta^{(\star)}_0$ &  $1 -i \, \frac{\omega}{4\pi T}$ & $ -i \, \frac{\omega}{4\pi T}$   \\
$\eta^{(\star)}_{r}$ &  $1 + i \, \frac{\omega}{4\pi T}$ & $i \, \frac{\omega}{4\pi T}$  
\end{tabular}
\end{center}
\caption{\em{The four possibilities for the index on the horizon, $\eta^{(lim)}$, in the limit of vanishing electric field.}}\label{indicestab}
\end{table}
From table \ref{indicestab}  we can see that to perform our study with a consistent ${E_p}\rightarrow 0$ limit we have to take the horizon index $\eta^{(h)}_-$ and at the singular shell $\eta^{(\star)}_0$.

As a further check that we chose the correct index at the singular shell we will show that our choice is the unique one that recovers the conductivity (\ref{jpofEp}) via the Kubo formula (\ref{kubo}).

\subsubsection{Fluctuations in the hydrodynamic approximation}

Here we are interested in solving equation (\ref{finiteEfluc}) in the limit of vanishing momentum. It is a consistency check that the solutions are regular both at the horizon and at the singular shell.

In the hydrodynamic limit we can expand the transverse fluctuation in a series in small $\omega$ and $q$. In order to perform this expansion we set $(\omega, q) \to \lambda_{hyd} (\omega, q)$ and perform the expansion for small $\lambda_{hyd}$. The series for the fluctuation then reads
 \be
 \dA_\perp=\dA_\perp^{(0)} + \lambda_{hyd}\, \dA_\perp^{(1)} + \OO(\lambda_{hyd}^2)\, .
 \ee
With the indices chosen in the last section we can write down the regular functions in the bulk order by order. At leading order in the hydrodynamic limit we calculate the equation of motion for the transverse gauge field fluctuation (\ref{finiteEfluc}) and find
\be
\left( e^{-\phi} \sqrt{-\gamma} \gamma^{\pperp} \gamma^{rr} {\dA^{(0)}_\perp}'  \right)'   =  0\, . \ee
This has a closed solution given by
\be
\dA_\perp^{(0)}(r) = C_0 + C_1 \int_{r_h}^r \frac{\dd \tilde r}{e^{-\phi} \sqrt{-\gamma} \gamma^{\pperp} \gamma^{rr}}\, .
\ee
where the integral is taken from the horizon out to a radius $r$. This integral is divergent at the singular shell and so the only regular solution is the the constant one $\dA_\perp^{(0)} (r)= C_0$.

Calculating the equation for the fluctuation at next to leading order we obtain the following equation
\be
\left( e^{-\phi} \sqrt{-\gamma} \gamma^{\pperp} \gamma^{rr} {\dA^{(1)}_\perp}'  \right)'  - i \omega \left( e^{-\phi} \sqrt{-\gamma} \gamma^{\pperp} \gamma^{(0r)} C_0  \right)'  = 0\, ,
\ee
which can be integrated twice to give
\be
\dA_\perp^{(1)}(r) = \int_{r_h}^r \frac{C_1 + i \omega e^{-\phi} \sqrt{-\gamma} \gamma^{\pperp} \gamma^{(0r)} C_0}{e^{-\phi} \sqrt{-\gamma} \gamma^{\pperp} \gamma^{rr}} \dd \tilde r +C_2\, .
\ee
This expression is completely regular at the horizon but needs to be regularized at the singular shell. With this regularity condition we obtain
\be
C_1 = - i \omega \frac{j_p C_0}{ \NN {E_p}} \, .
\ee

We now calculate the conductivity from the Kubo relation (\ref{kubo}), which relates the transport coefficient to the transverse correlation function. Using this prescription we obtain:
\be\label{anconduc}
\sigma =  -\NN \lim_{\omega\to0} \im  \frac{C_1 + i \omega e^{-\phi} \sqrt{-\gamma} \gamma^{\pperp} \gamma^{(0r)} C_0}{\omega  C_0}\Bigg|_{r_{bou}} \!\!\!\!\!\! =  \frac{j_p}{{E_p}} \left( 1+ \frac{(2\pi \alpha')^2 {E_p}^2}{g_{00} g_{ii} }\right) \Bigg|_{r_{bou}}\!\!\!\!\!\! ,
\ee
The second term vanishes when evaluated at the boundary and we recover the result of  \cite{Karch:2007pd}.
This is a good clue that we have chosen the correct index at the singular shell. Had we chosen the index $\eta_r^{(\star)}$ we would have found a different answer (incompatible with the result of the macroscopic conductivity calculation), which in the limit of vanishing electric field reads
$\sigma_{E_p=0} = -\lim_{E_p\to 0}j_p / E_p$,
exhibiting a change of sign that can be traced back to the time reversal that also appears in table \ref{indicestab}. With this positive result in hand we can now go on to calculate the spectral function away from the hydrodynamic limit which will ultimately give us the photoproduction of the strongly coupled plasma in the presence of an external electric field.

\section{Fluctuations in the  $D3/D7$ system}\label{numerical}

Here we specialise once again to the $D3/D7$ system and turn to numerical calculations of physically relevant quantities. Having shown that our choice of indices is the unique one to recover the conductivity while also recovering the correct $E_p \to 0$ limit (the usual incoming wave boundary conditions) we are now able to study the spectral function for mesons dissociated both by the electric field and by finite temperature effects. 

Using the radial coordinate introduced in section \ref{specd3d7}, the gauge field can be expanded in the UV (small $u$ region) in terms of the two independent solutions 
\be
\dA_\perp = \dB_1 \dA^{(1)} (u) + \dB_2 \dA^{(2)}(u)\, ,
\ee
where
\beqa
\dA^{(1)} (u) & = & 1 + \sum_{i=2}^\infty a^{(1)}_i u^i + h\, \dA^{(2)} \log u\, , \\
\dA^{(2)} (u) & = & u + \sum_{i=2}^\infty a^{(2)}_i u^i\, .
\eeqa
All the coefficients $a^{(n)}_i$ and $h$ are determined by recursion relations\footnote{for example $h= \qn^2 - \wn^2$ with  $\wn \equiv \omega/(2\pi T)$ and $\qn \equiv q/(2\pi T)$.}, where $\dB_m$ are the two connection coefficients. Applying the lorentzian AdS/CFT prescription (eq. (\ref{twopoint})) we find for the two-point function
\be
\Pi^\perp = \frac{N_c N_f T^2}{2} \lim_{u\to 0} \frac{{\dA_\perp}'(u)}{\dA_\perp(u)} = \frac{N_c N_f T^2}{2} \left( \frac{\dB_2}{\dB_1} + h \left( 1 + \log(u\to 0)\right)\right)\, ,
\ee
leading to an expression for the transverse part of the spectral function
\be
\chi^\perp= -2 \im \Pi^\perp = -   N_c N_f T^2  \im  \frac{\dB_2}{\dB_1} \equiv    N_c N_f T^2  \tilde \chi^\perp\, .
\ee
This of course will be a function of $\efield, \bardens$ and $m_q$ leading to a large parameter space. We split this analysis into two separate areas: The first is confined to solutions with zero spatial momentum which will allow us to see the appearance of quasiparticle like peaks in certain limits of the parameter space, and also to compute the conductivity numerically as the slope 
$\sigma = 1/2\lim_{\omega\to 0}\chi^\perp(\omega,0)/\omega$.
The second will be solutions with lightlike spatial momentum in which we will be able to calculate the photoproduction rate.

To perform the study we will numerically integrate equation (\ref{finiteEfluc}) from the singular shell to the boundary. As mentioned in the previous section, we will use a regularising index $\eta_0^{(*)}=0$, so that the function $\dA_\perp(u)$ is regular at the singular shell. Once the numerical solution is obtained we read the asymptotic behavior near the boundary and from that data we extract the coefficients $\dB_m$, which will depend on all the parameters considered.

\subsection{Transverse fluctuations at zero momentum}
Setting $\qn=0$ allows us to plot the spectral function as a function of $\wn$  and examine the evolution of it across the parameter space spanned by $(\efield, \bardens, m_q)$.  A  fairly important change in shape is observed upon crossing the conical embedding
transition region in the $( m_q, \efield)$ plane (see fig. \ref{transregion}). 
\begin{figure}[ht]
\begin{center}
\includegraphics[scale=1.45]{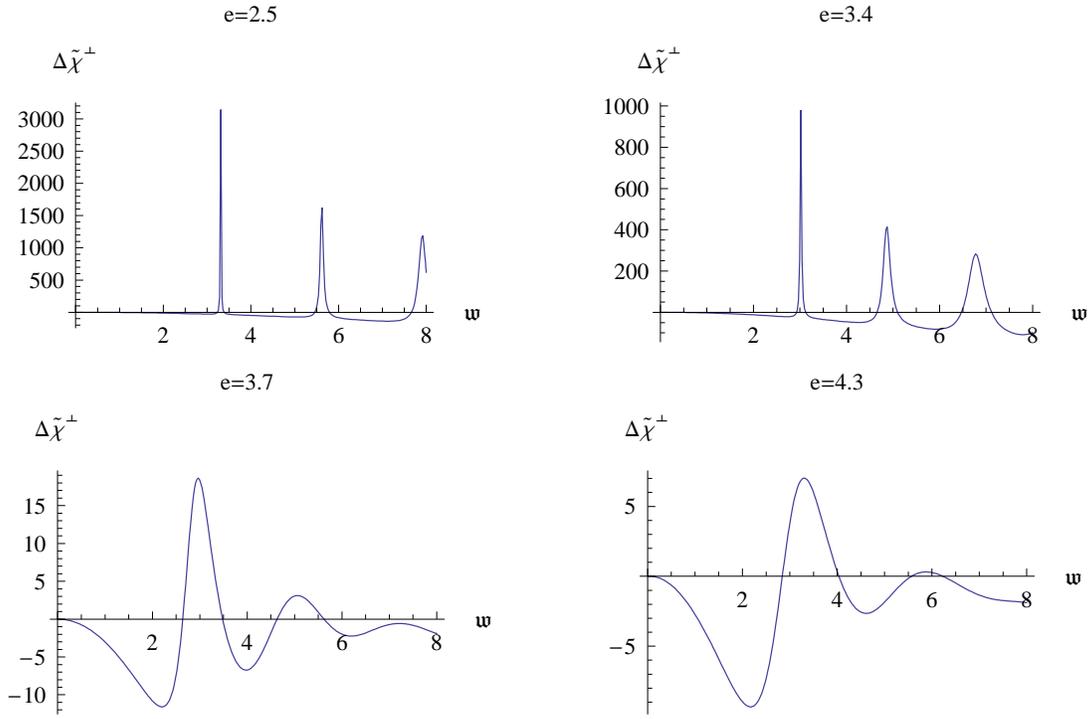}
\caption{\em \label{spectraltransition0012}
The difference, $\Delta\tilde{\chi}^\perp$, between the spectral function $\tilde{\chi}^\perp(\wn)$ and the zero temperature limit with increasing values of the electric field $\efield$ at fixed value of the baryon density ($\bardens=0.012$) and mass ($m_q=3.5$). After crossing the line of conical embeddings the peaks broaden very quickly into wide oscillations which cannot be interpreted as quasiparticles.}
\end{center}
\end{figure}
In figure  \ref{spectraltransition0012} we show an example sequence of spectral functions for a range of values of $\efield$ and with $\bardens = 0.012$ and $m_q = 3.5$. Over a very small range, $3.4\lesssim \efield \lesssim 3.7$, the magnitude of the spectral function drops by around two orders of magnitude and the sharp peaks disappear completely, leaving a smooth profile with very wide oscillations which can no longer be interpreted as quasiparticles. This rapid crossover in behavior is a relic of the very fast change in behavior of the embedding in a small region of the $(m_q,\efield)$ parameter space.

A physical observable related to the spectral function at zero momentum is the conductivity. 
A perfect agreement between our numerical computation of this quantity and that performed analytically, eq. (\ref{anconduc}), is found
\be\label{eqcond}
\dlcond= \frac{1}{2} \lim_{\wn \to 0} \frac{\tilde \chi^\perp}{\wn}= \sqrt{\sqrt{1+{\efield}^2}\left(1-\psi_\star^2\right)^3+\frac{\bardens^2}{1+{\efield}^2}}\,  .
\ee

\subsection{Lightlike momentum and photoproduction}

Here we generalise the previous calculation to the case of lightlike momenta. This allows us to study photoproduction and discover how an electric field affects the brightness of the plasma \cite{hep-th/0607237,arXiv:0709.2168}.

\begin{figure}[ht]
\begin{center}
\includegraphics[scale=1.2]{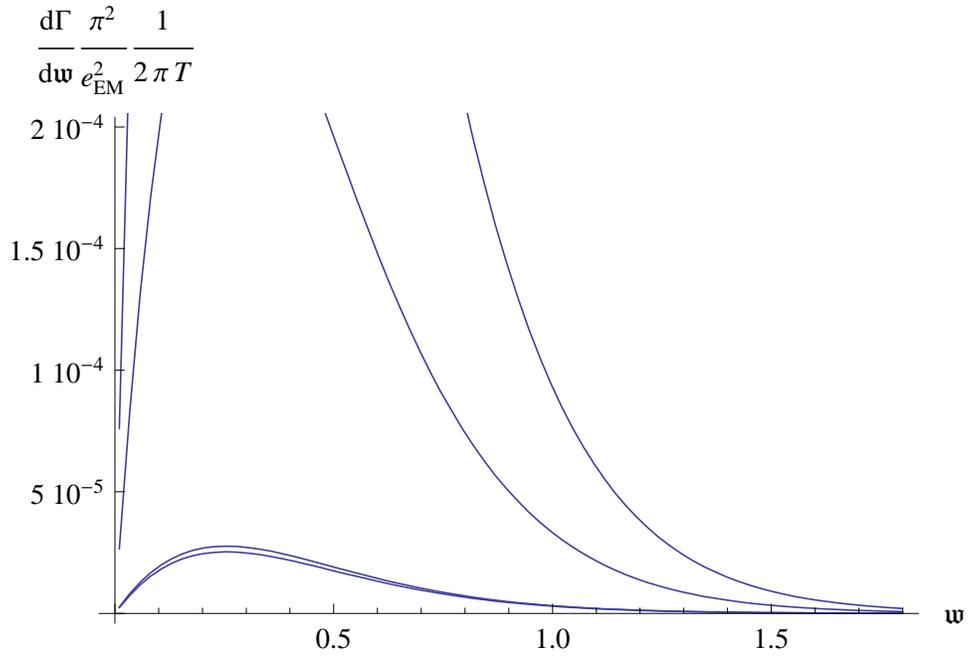}
\caption{
\em \label{photoplotcurve}
Photoproduction density rate in the direction of the electric field, from low to high, at values of $\efield=2.3,\,  2.5,\, 2.7$ and $3$ with $\bardens=0.05$. 
 }
\end{center}
\end{figure}

The emission rate for real photons is controlled by the spectral function evaluated at lightlike momenta $\omega = |{\bf k}|$. Given that on the light cone $\Pi_{||}=0$ (otherwise we would have a divergence in the retarded Green's function), we can express the emission rate entirely in terms of $\Pi_\perp$ as follows
\be
d\Gamma_\gamma =-\left. \frac{d{\bf k}^3}{(\pi)^3}\frac{e_{EM}^2}{2 | {\bf k} |} \frac{\im \Pi_\perp(k^\mu)}{e^{\omega/ T}-1}\right\vert_{\omega = | {\bf k} |}\, ,
\label{difphoto}
\ee
where $e_{EM}$ is the electromagnetic coupling constant. Since our system lacks isotropy, strictly speaking, the above expression only  applies to  photons with $k^\mu= \omega(1, 0,0,1)$ which propagate parallel to the electric field $\vec E= (0,0,E_z)$\footnote{we would like to thank David Mateos for pointing out this subtlety}.
In figure \ref{photoplotcurve} we see a series of curves that represent the value of 
$d\Gamma(\wn)/d\wn$ normalized by $(2\pi T) e_{EM}^2/\pi^2$ . The Boltzman factor means that the relevant structure can only be seen in the range of frequencies $\wn\in(0,2)$. The curves are plotted for $m_q=3$ and increasing values of the electric 
field $\efield = 2.3, 2.5, 2.7$ and $3$. A strong enhancement is observed upon traversing the crossover region.

\begin{figure}[ht]
\begin{center}
\includegraphics[scale=0.55]{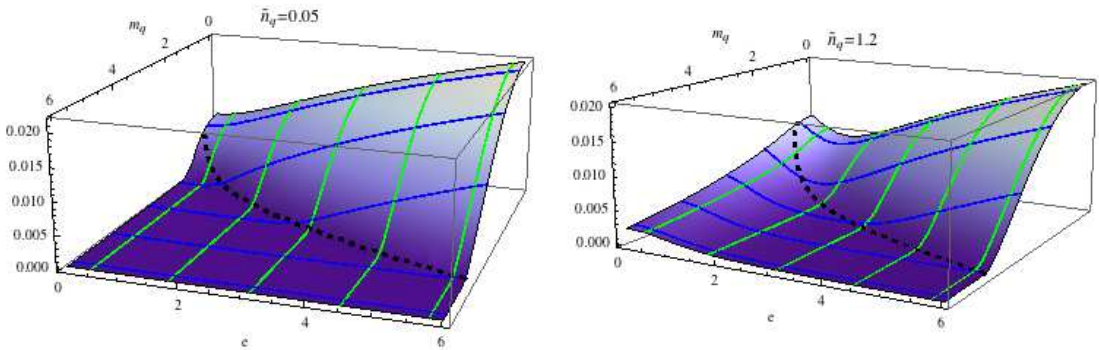}
\caption{
\em \label{integratedphotoplot}
Integrated photoproduction $(2\pi)^3\Gamma/e_{EM}^2 $  as a function of $(m_q,\efield)$.  }
\end{center}
\end{figure}

The rates are  similar and exhibit a peak at $\wn \sim 0.25$.
We could choose this as a representative value of the overall magnitude of the photoproduction, and see how it changes as
we move across the $(m_q,\efield)$ space. Alternatively we can integrate (\ref{difphoto}) in $\omega\in (0,\infty)$, which gives
the  total number density of thermal photons emitted parallel to the electric field (per unit momentum space solid angle). The result  can be seen in  figure \ref{integratedphotoplot}. A remarkable similarity with the plots in figure \ref{conductplot} is observed.
Of course a knowledge of the  differential photoproduction rate for $\vec q$ not parallel to  $\vec E$ would be extremely interesting. We leave this analysis for a further study.

\section{Discussion and conclusions}\label{conclusions}

In this paper we have studied in detail the effect of an external electric field on the physical properties of a large $N_c$ finite temperature Yang Mills plasma with a single flavour in the quenched approximation.  In addition to the strength of the electric field the free parameters of our model have been the baryon density and the quark mass and we have explored this three-dimensional parameter space thoroughly. The singular shell which is a direct consequence of a large enough ratio of the electric field strength to the quark mass appears to play a key role in the quantities of the gauge theory living on the AdS boundary. Although the singular shell does not have the geometrical properties to label it a horizon, in many ways its effects appear to be similar to those of a real black hole horizon. In particular the attractor-like behavior whereby the solution of the embedding is fixed uniquely by a single boundary condition constrains the solutions greatly. Continuing the embedding inside the singular shell gives us solutions which may or may not have a conical singularity, depending on the values of $e$ and $m_q$ for a particular value of $\bardens$. Indeed this conical singularity appears and then disappears as we traverse the $(e,m_q)$ parameter space. The region in which this transition happens appears to play a key role as a crossover region in the behavior of many physical quantities. However, it may be interpreted that this region of conical singularities is only a symptom of what is happening at the singular shell.

In \cite{arXiv:0805.2601} it was argued that the size of the effective horizon induced on a probe $D7$-brane greatly affected the spectral function, with a small induced horizon leading to a series of sharp peaks in the spectral function and long-lived quasiparticles, whereas a large effective horizon led to a smooth spectral function and the disappearance of the quasiparticle interpretation as the poles in the quasinormal mode spectrum disappeared into the complex frequency plane. This phenomenology is related to the fact that the incoming wave boundary conditions at the horizon allow the absorption of energy into the system more rapidly if the effective horizon is larger. Quasiparticles have shorter lifetimes if, in the holographic picture, their energy can be dissipated into a larger effective horizon.

Although the interpretation of the singular shell as any sort of effective horizon is unclear, we do appear to have a very similar phenomenology in the work presented here. For a large induced singular shell area on the $D7$-brane (corresponding to a small value of $\psi_\star$) the spectral function is smooth, while for a small induced singular shell area, the spectral function is sharply peaked. The crossover region is then traced to a small region in parameter space where the area of the induced singular shell changes very quickly from a small value to a large value, allowing, for instance, for several orders of magnitude change in the photoproduction rate.

One may ask if this is not due to the effective horizon itself, however it appears that this is not so. As we pass through the cross-over region, the size of the effective horizon does not increase monotonically, whereas it seems that the widths of the peaks in the spectral function, for example, do. If the widths of peaks in the spectral function were determined by the size  of the effective horizon as in \cite{arXiv:0805.2601} one should see the same monotonic behaviour. For this reason it appears that the important IR quantity is the area of the effective singular shell, where again the incoming wave boundary conditions which completely fix the UV behaviour, seem to have a similar effect to the horizon in the $e=0$ case. In addition, because of the attractor-like mechanism at the singular shell, effectively the behavior outside the singular shell is completely shielded from what is happening inside. Any additional force that one applied to the brane inside the singular shell will not be felt outside. It is however true that the changes at the singular shell are reflected both in the UV physics and the behavior of the brane within the singular shell. For this reason the line of conical embeddings simply appears to be a good marker of the region where we have discovered such interesting effects.

It should be noted that the calculation for the susceptibility, eq (\ref{eq.susc1}), does appear to rely on the behavior inside the singular shell. This is because we have imposed the vanishing of the time component of the gauge field on the horizon. A check that the susceptibility does not in fact depend on the region inside the singular shell would be to calculate it via the Einstein relation for which we would need to calculate the diffusion constant from the longitudinal correlator. We leave this investigation for the future.

There may be hints to the physics behind the crossover region coming from the analytic expression from the conductivity. Indeed we can see that there are two terms in this expression, one of which is clearly related to the size of the induced singular shell area and increases with increasing electric field strength, while the other decreases with the electric field strength. As you increase $\efield$ or $1-\psi_\star^2$ the dominant term in the conductivity will change from one to the other. Because $m_q$ and not $\psi_\star$ is the more physical parameter in our gauge theory the contribution of the first term is less clear when we plot physical quantities as a function of $m_q$. We conjecture that the two competing effects which are very clear in the expression for the conductivity (namely pair-production and free charges respectively) may explain the rapid change in behavior across the crossover region. It would certainly be interesting to investigate the interplay between the two competing effects in more depth.

It would be extremely exciting if one could test these predictions of a fast crossover region in the laboratory. However, it appears that the current technical limitations in heavy-ion experiments do not allow for the investigation of the quark gluon plasma in the presence of external electric fields. What may be more practical would be a condensed matter system modeled holographically, and demonstrating a similar phenomenology to the one studied in this paper. An obvious example would be the defect $D3/D5$ system.

\acknowledgments
We would like to express our gratitude to  Daniel Arean, Andy O'Bannon, Veselin Filev,
David Mateos, Rene Meyer, Alfonso Ramallo and Dimitrios Zoakos. 
This  work was supported in part by by the MICINN and FEDER (grant FPA2008-01838), by the Spanish Consolider-Ingenio 2010 Programme CPAN (CSD2007-00042), by Xunta de Galicia (Conselleria de Educacion and grant PGIDIT06 PXIB206185PR).
J.T. and J.S. have been supported by MICINN of Spain under a grant of the FPU program and
 by the Juan de la Cierva program respectively.

\end{document}